\documentclass[12pt,hidelinks]{article}
\usepackage[body={17.5cm, 22cm},right=2cm]{geometry}
\usepackage[utf8]{inputenc}
\usepackage{color}
\usepackage{amssymb,graphicx}
\usepackage{amsmath}
\usepackage{epsfig}
\usepackage{lmodern}
\usepackage[normalem]{ulem}
\usepackage{bm}
\usepackage[bookmarks=true,pdfborder={0 0 0}]{hyperref} 
\usepackage[
    natbib=true,
    style=numeric-comp,
    sorting=none
]{biblatex}
\usepackage{booktabs}   %
\usepackage{siunitx}    %
\usepackage{orcidlink}
\usepackage{graphicx}   %
\usepackage{subcaption}
\usepackage{float}
\allowdisplaybreaks

\DeclareUnicodeCharacter{223C}{\ensuremath{\sim}}   %
\DeclareUnicodeCharacter{2212}{\ensuremath{-}}      %
\DeclareUnicodeCharacter{2243}{\ensuremath{\simeq}} %
\DeclareUnicodeCharacter{2248}{\ensuremath{\approx}}%
\DeclareUnicodeCharacter{00B1}{\ensuremath{\pm}}    %
\DeclareUnicodeCharacter{2013}{--}                  %
\DeclareUnicodeCharacter{2014}{---}    
\DeclareUnicodeCharacter{0308}{\"{}}

\newcommand{\GeV}{{\, {\rm GeV}}}

\definecolor{myorange}{RGB}{199,146,32}
\definecolor{myblue}{RGB}{35,20,180}

\begin{document}
\setcounter{page}{0}
\thispagestyle{empty}

\parskip 3pt

\font\mini=cmr10 at 2pt

\begin{titlepage}
\noindent \makebox[14.8cm][l]{\footnotesize \hspace*{-.2cm} }{\footnotesize 
DESY-26-107}  \\  [-1mm]
~\vspace{1cm}
\begin{center}

{\Large \bf Imperfect axions with no domain wall problem}

\vspace{0.8cm}

		{\large
        Marco Gorghetto\,\orcidlink{0000-0002-5479-485X},$^a$\,
          Edward Hardy\,\orcidlink{0000-0003-3263-6575},$^b$\,
            Gilad Perez\,\orcidlink{0000-0002-3878-1821},$^c$\,
            and Stefan Stelzl\,\orcidlink{0000-0001-5964-1054},$^d$\,
          }

		\vspace{.8cm}

		{\normalsize { \sl $^{a}$ 
				Deutsches Elektronen-Synchrotron DESY, Notkestr. 85, 22607 Hamburg, Germany

                }}

            \vspace{.3cm}
		{\normalsize { \sl $^{b}$          Rudolf Peierls Centre for Theoretical Physics, University of Oxford, \\ Parks Road, Oxford OX1 3PU, UK
				}}

    \vspace{.3cm}
		{\normalsize { \sl $^{c}$ 
			        Department of Particle Physics and Astrophysics, Weizmann Institute of Science,\\
				Herzl St 234, Rehovot 761001, Israel
                }}

    \vspace{.3cm}

		{\normalsize { \sl $^{d}$ 
			Institut de F\'{i}sica d'Altes Energies (IFAE) and
Barcelona Institute of Science and Technology (BIST), 
Campus UAB, 08193 Bellaterra (Barcelona), Spain
                
                }}

\end{center}

\vspace{1.0cm}
\begin{abstract}
\vspace{0.2mm}

Post-inflationary axion theories with domain wall number $N>1$ suffer from a domain wall problem. We show analytically, and confirm via simulations, that there is a region of phenomenologically viable parameter space for such theories with additional PQ-violating operators. Contrary to the conventional picture, in such theories the beyond-QCD PQ violation destroys the axion strings \emph{before} the QCD axion potential becomes cosmologically relevant. This regime is realized for axion decay constants $f_a \lesssim 10^{10} \, \rm GeV$, pointing to a QCD axion mass in the $10^{−3}-10^{−2}\,{\rm eV}$ range. The required additional PQ violation induces a neutron electric dipole moment within two orders of magnitude of the current experimental limit, placing it within the reach of upcoming neutron electric dipole moment experiments.

\vspace{1cm}

\end{abstract}

\end{titlepage}

\section{Introduction}
The QCD axion provides an elegant solution to the strong CP problem~\cite{Peccei:1977hh,Peccei:1977ur,Weinberg:1977ma,Wilczek:1977pj}, and can also account for the observed dark matter abundance~\cite{Abbott:1982af,Dine:1982ah,Preskill:1982cy,Marsh:2015xka,Adams:2022pbo}. In 4-dimensional realizations, it arises as a pseudo-Nambu--Goldstone boson of a spontaneously broken global $U(1)$ Peccei--Quinn (PQ) symmetry. 

The cosmology of the axion depends on whether the PQ symmetry is spontaneously broken before or after inflation. In the former, preinflationary, case, a relic abundance of axions is produced via the misalignment mechanism. The absence of isocurvature perturbations in CMB observations \cite{Planck:2018jri} places strong bounds on the inflationary Hubble scale in this scenario~\cite{Steinhardt:1983ia,Seckel:1985tj,Linde:1985yf,Graham:2025iwx}, making it difficult to reconcile with high-scale inflation.

On the other hand, in the post-inflationary scenario (which can be realized also in string theory~\cite{Petrossian-Byrne:2025mto,Petrossian-Byrne:2026zwl,Loladze:2025uvf}) the PQ symmetry is broken after inflation and there is negligible isocurvature on CMB scales%
. In this case, spatial inhomogeneities in the PQ-breaking field give rise to axion strings via the Kibble mechanism \cite{Kibble:1976sj}. These strings evolve toward a scaling solution and produce axions that can account for all of dark matter~\cite{Sikivie:1982qv,Vilenkin:1982ks,Vilenkin:1984ib,Davis:1986xc,Harari:1987ht,Gorghetto:2020qws,Saikawa:2024bta,Buschmann:2021sdq,Benabou:2024msj,Kim:2024wku,Gherghetta:2025fip,Wantz:2009it}. As QCD becomes strongly coupled, it generates a potential for the axion whose magnitude grows rapidly as the Universe cools, resulting in the strings becoming attached to domain walls (DWs). The DW number $N\geq1$ depends on the axion UV completion. For $N=1$, the minimum of the axion potential is unique and the string-wall network is unstable, and it is expected to annihilate quickly~\cite{PhysRevLett.48.1867}. The case $N>1$, motivated by, e.g., DFSZ or composite axion models~\cite{Randall:1992ut,Azatov:2025mep,Redi:2016esr,Dine:1981rt,Zhitnitsky:1980tq}, and, more generally, in models with QCD-matter states beyond a single vector-like pair, leads instead to a stable network of DWs interpolating between the $N$ inequivalent minima, which would overclose the universe -- the DW problem~\cite{Zeldovich:1974uw}. 

A viable post-inflationary $N>1$ cosmology therefore requires an additional PQ-breaking potential that selects a unique global minimum so that the DWs decay~\cite{Zeldovich:1974uw,Sikivie:1982qv,Gelmini:1988sf,Gelmini:2022nim}. Moreover, the DW decay must occur early enough that the resulting axions do not overproduce dark matter~\cite{Hiramatsu:2012gg,Hiramatsu:2013qaa,Kawasaki:2014sqa}. However, barring a tuning between the alignment of the QCD potential and the PQ-breaking one, this additional potential is strongly constrained since it induces a non-zero $\theta$ and thus a neutron electric dipole moment (nEDM). Current nEDM measurements imply $\theta \lesssim 10^{-10}$, with sizable improvements expected in the near future~\cite{Alarcon:2022ero}. See e.g. Refs.~\cite{Lazarides:1982tw,Gherghetta:2025fip,Lu:2023ayc,Choi:2026oqz,Ibe:2019yew,Lee:2025zpn,Larsson:1996sp,Dine:2023qsq,Hor:2026dlb,Zhang:2023gfu} for other approaches to the DW problem and related discussion.

The conventional picture of the dynamics in the presence of PQ breaking is as follows~\cite{Sikivie:2006ni}. %
First, a network of stable DWs forms from the QCD potential at the temperature $T_\star$, when the Hubble parameter $H$ equals the axion mass $m_a$. This network can be destroyed by the bias term induced by the PQ-violating operator only at later times, once the axion mass approaches its vacuum value near the QCD crossover temperature, $T_c \simeq 150$\,MeV. The decay of this long-lived network occurs once the energy-density difference $\Delta V$ between axion minima satisfies $\Delta V \sim \sigma H$, where $\sigma \sim m_a f_a^2$ is the DW tension. Even taking $\theta$ as large as allowed by  nEDM searches, the axions produced from DW decay would be on the edge of overproducing dark matter for values of the axion decay constant $f_a$ consistent with astrophysical bounds~\cite{Kawasaki:2014sqa,Beyer:2022ywc,Ringwald:2015dsf}.

In this work we show that, if the PQ-breaking operators are within two orders of magnitude of the maximal size allowed by nEDM measurements and $f_a$ is sufficiently small (e.g.\ $f_a\lesssim 10^{10}$\,GeV for maximal PQ breaking), the conventional dynamics just described do not occur. Instead, the PQ-breaking operators become cosmologically relevant \emph{before} the QCD potential is generated at $T_\star$. In that case, the strings undergo a qualitatively different evolution: the network is significantly disrupted by the PQ-breaking term before the QCD DW network ever forms at $T_\star$, and annihilates around this earlier time. %
As we will see, this opens up a region of phenomenologically viable parameter space in which the DW problem is solved consistently with the observed dark matter abundance and astrophysical bounds. %
Specifically, this is the case if the PQ-breaking potential has a unique global minimum and the DW number is not too large: $N=6$, as motivated by the DFSZ model, can still realize this new regime, but much larger values cannot.%

Our study is limited by the modest hierarchy between the string-core scale and the PQ-breaking mass accessible to simulations, and by the possible backreaction on wall decay of the axions abundantly emitted during scaling. We leave these points to future work, where they might be addressed by dedicated simulations.

\section{PQ-breaking operators and axion quality}

To illustrate our main point, we consider the Lagrangian for the complex scalar field $\phi(x)$,
\begin{equation}\label{eq:Lag}
    \mathcal{L} = |\partial_\mu \phi|^2 - \frac{m_r^2}{2 v^2} \left(|\phi|^2-\frac{v^2}{2}\right)^2 + b\frac{|\phi|^n}{M_p^{n-3}} \phi + h.c. \, ,
\end{equation}
where the second term drives spontaneous breaking of the PQ symmetry, the last term its explicit breaking, and $M_p$ is the Planck mass. The axion $a(x)$ is the resulting pseudo-Goldstone boson, defined as the phase of $\phi= \frac{1}{\sqrt{2}}\left(v+r\right) e^{i a/v}$, while the radial mode $r(x)$ is a heavier field of mass $m_r$. Here $n+1$ is the dimension of the leading PQ-breaking operator. We use $n=0$ in the simulations below, but the phenomenology we describe holds equally for $n>0$. 
We focus on the case of a unique minimum of the PQ-breaking potential without any metastable minima, as occurs in Eq.\,\eqref{eq:Lag} (in which the PQ-breaking operator carries unit PQ charge). At the end of Sec.~\ref{ss:PQbefore} we argue that a similar phenomenology can still occur if metastable minima are present in the PQ-breaking potential. There we also comment on the case that the PQ-breaking potential has degenerate minima.

The complex scalar also couples to $N$ fermions charged under PQ and QCD. After an anomalous chiral rotation, the axion-gluon coupling $N \frac{a}{v} \frac{\alpha_s}{8\pi}G_{\mu\nu}^a \tilde{G}^{\mu \nu a}$ is generated in the IR, where $\alpha_s$ is the strong fine-structure constant and $G_{\mu\nu}$ $(\tilde{G}^{\mu \nu})$ is the gluon (dual) field strength. The axion decay constant can then be expressed as $f_a = v/N$. Non-perturbative dynamics induce, via this coupling, the QCD-generated part of the axion potential $V_{\rm QCD}(a)$, minimized at the CP-conserving value $a/f_a=0$, with a  periodicity of $2\pi f_a$, and yielding, in the minimal theory, the zero-temperature mass $m_a = 5.69(5)\,\mu{\rm eV}\,(10^{12}\,{\rm GeV}/f_a)$~\cite{DiVecchia:1980yfw, GrillidiCortona:2015jxo,Weinberg:1977ma,Gorghetto:2018ocs}. %

In the vacuum, the full axion potential is
\begin{equation}\label{eq:QCDpotential}
V(a) = V_{\rm QCD}(a)  - m_{\rm PQ}^2 v^2 \cos\left(\frac{a}{v}-\delta\right) \, \equiv \, V_{\rm QCD}(a)+V_{\rm PQ}(a)\, ,
\end{equation}
where $m_{\rm PQ}^2 \equiv  |b|\frac{(v/\sqrt2)^{n-1}}{M_p^{n-3}}$, and $\delta$, expected to be $\mathcal{O}(1)$, measures the misalignment between the minimum of the axion potential from QCD and the phase of the PQ-breaking coefficient $b$. %
In phenomenologically viable theories, at zero temperature the PQ-breaking mass $m_{\rm PQ}$ must be much smaller than $m_a$, as it shifts the minimum of the potential away from  the CP-conserving point, inducing an effective strong CP phase
\begin{equation}\label{eq:theta-eff}
    \theta_{\rm eff} \equiv \frac{\langle a\rangle}{f_a}=  N \sin\delta\,\frac{ m_{\rm PQ}^2}{m_a^2} \,  .
\end{equation}
Together with the non-observation of a neutron EDM~\cite{Abel:2020pzs}, this limits the maximal size of $m_{\rm PQ}$ to %
\begin{equation}
\label{eq:maxPQviolation}
    m_{\rm PQ}^2 \lesssim  10^{-10} \frac{m_a^2}{N\sin \delta}\,.
\end{equation}
The generation of a non-zero $\theta_{\rm eff}$ from PQ-violating operators -- even if Planck-suppressed -- is known as the axion quality problem, see e.g.~\cite{Georgi:1981pu,Kamionkowski:1992mf,Barr:1992qq,Holman:1992us}. In this work we do not address this, and simply assume that the leading PQ-breaking term is small enough to not spoil the solution to the strong CP problem, according to Eq.\,\eqref{eq:maxPQviolation}, but still being sufficiently large to have important cosmological implications.%
\footnote{It is plausible that the leading PQ-breaking operator is exponentially suppressed rather than suppressed by a large power $n\gg1$, but still temperature-independent, in which case the phenomenology below would be unchanged.%
}

\section{Early destruction of the $N>1$ domain walls}

The bound on the PQ-breaking term in Eq.\,\eqref{eq:maxPQviolation} holds at zero temperature. A key point is that the axion mass $m_a(T)$ induced by QCD is %
strongly suppressed at $T\gg T_c\simeq 150$\,MeV. As we now show, the temperature-independent PQ-breaking potential $V_{\rm PQ}$ in Eq.\,\eqref{eq:QCDpotential} may dominate over $V_{\rm QCD}$ before the QCD crossover phase transition, leading to an early decay of the $N>1$ DW networks.

In the post-inflationary scenario, axion strings form at the PQ phase transition and evolve into an attractor scaling regime, in which the number of strings per Hubble patch $\xi\simeq 0.24\log(m_r/H)$ remains constant up to slow logarithmic running~\cite{Gorghetto:2018myk,Gorghetto:2022ikz,Buschmann:2021sdq,Kim:2024wku,Saikawa:2024bta,Benabou:2024msj}; here $H=1/(2t)$ is the Hubble parameter during radiation domination. During scaling, the axions radiated from strings contribute to the relic abundance, and may account for all of it depending on $f_a$. The string network exits the scaling regime once the axion potential of Eq.\,\eqref{eq:QCDpotential} becomes cosmologically relevant, at temperatures parametrically above $T_c$ (typically at $T_\star \sim 20\,T_c$, see Eq.\,\eqref{eq:Tstar} in App.~\ref{app:estimates}). %

At $T\gg T_c$ the QCD potential is well approximated by $V_{\rm QCD}(a)=-m_a^2(T)\,v^2/N^2\cos(Na/v)$. Around a string, $a(x)$ nontrivially wraps $(-\pi v,\pi v]$. Thus, for $N=1$, once $V_{\rm QCD}$ becomes important, a single axion DW is formed around a string, which annihilates the string network. For $N>1$, the axion explores multiple degenerate minima of $V_{\rm QCD}$, corresponding to $N$ stable DWs attached to the string, each interpolating between two of the $N$ inequivalent minima, which pull it in different directions with equal strength. The conventional picture is that %
the PQ-breaking term $V_{\rm PQ}$ in Eq.\,\eqref{eq:QCDpotential} is a small perturbation on $V_{\rm QCD}$ that lifts this degeneracy, biasing one DW to pull more strongly and causing the network to decay. %

\begin{figure}
    \centering
    \includegraphics[width=0.905\linewidth]{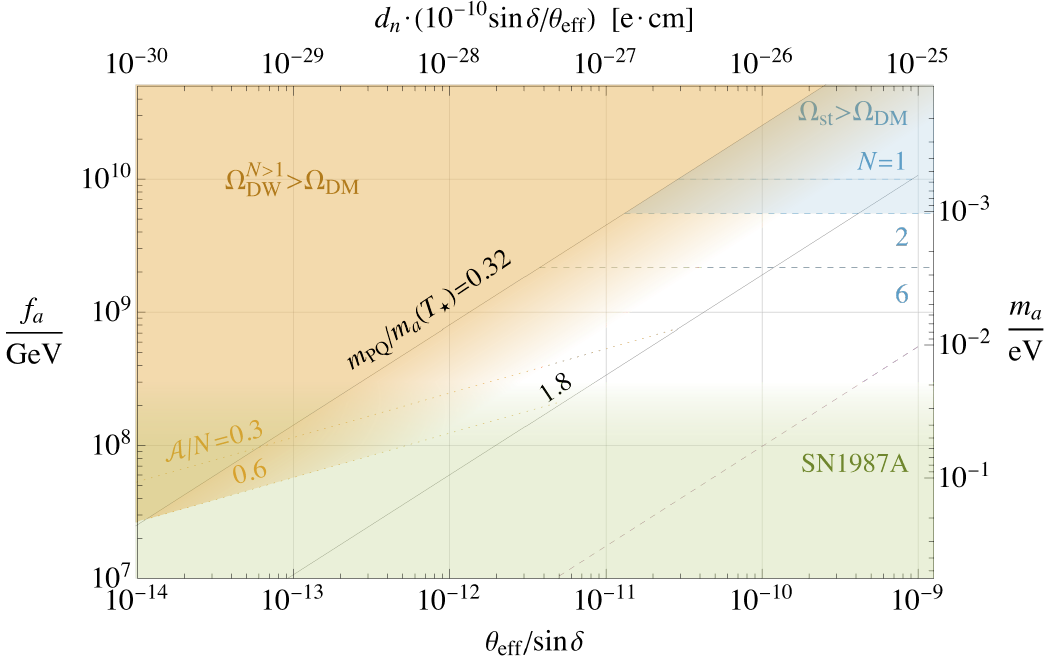}
    \caption{{\small Constraints on the axion decay constant $f_a$ and the PQ-breaking term, parametrized by $\theta_{\rm eff}$, for the QCD axion with $N>1$. The line $m_{\rm PQ}/m_a(T_\star)\sim 0.32$, shown for $N=2$, %
    separates the two regimes%
    : below it, PQ breaking becomes relevant before the QCD potential, and the $N>1$ QCD DW network never forms; above it, the long-lived QCD network forms and eventually decays due to the small PQ bias. Owing to the logarithmic enhancement of the string tension, we expect the actual transition between the two regimes to shift to $m_{\rm PQ}/m_a(T_\star)\sim1.8$. For larger $N$ these lines shift down by a factor $(N/2)^{3/4}$ (not shown). The blue region shows the bound from dark-matter overproduction by axions radiated during the string scaling regime; the orange region shows the bound from axions emitted by the long-lived DW network for $N=2$, and the band $\mathcal{A}/N=0.3$--$0.6$ reflects the uncertainty in the DW area per Hubble patch. We show the supernova bound in green (SN1987A). The bulk of the allowed parameter space lies in the region where the DWs disappear early due to the additional PQ breaking. The relativistic axions emitted by strings might inhibit DW formation: in this case  early decay induced by PQ-breaking only occurs below the purple dashed line, see App.~\ref{app:waves}. 
    }
    }
    \label{fig:results}
\end{figure}

The QCD potential has an explicit temperature (i.e., in the early Universe, time) dependence.  Meanwhile, we assume that the PQ-breaking contribution to the potential is generated by physics at high energy scales and is thus temperature-independent at the scales of interest.\footnote{Relaxing this assumption and allowing the PQ-violating term to be temperature-dependent, as would occur e.g. if it arises from a hidden sector that confines after the QCD phase transition, alters the phenomenology, and we do not discuss this possibility here.} For $T\gg T_c$, the dilute instanton gas approximation~\cite{Gross:1980br}, supported by lattice QCD calculations~\cite{Bonati:2015vqz,Petreczky:2016vrs,Borsanyi:2016ksw,Burger:2018fvb,Bonati:2018blm}, suggests that
\begin{equation}
    m_a^2(T) \simeq  m_a^2(0) (\Lambda/T)^\alpha \, ,
\end{equation}
with $\Lambda\simeq 150$ MeV, $\alpha\simeq 8$, while $m_a^2(T)$ saturates to $m_a^2(0)$ at $T\ll T_c$.   It is exactly this temperature dependence that makes the PQ-breaking potential important before the QCD one. %
 In particular, the latter becomes important at the temperature $T_\star$ when $m_a(T_\star) = H(T_\star)\equiv H_\star$. Owing to the strong temperature dependence, the QCD mass at this time is far smaller than when it has saturated at late times (see App.~\ref{app:estimates}):
\begin{equation}\label{eq:ma2soma2}
    \frac{m_a^2(T_\star)}{m^2_a(0)} \simeq 1.4\cdot 10^{-10}\left[\frac{f_a}{10^{10} \,{\rm GeV}}\left(\frac{g_{*}(T_\star)}{60}\right)^\frac12\right]^{\frac{4}{3}} \,.
\end{equation}
\begin{figure}
    \centering
    \includegraphics[width=0.58\linewidth]{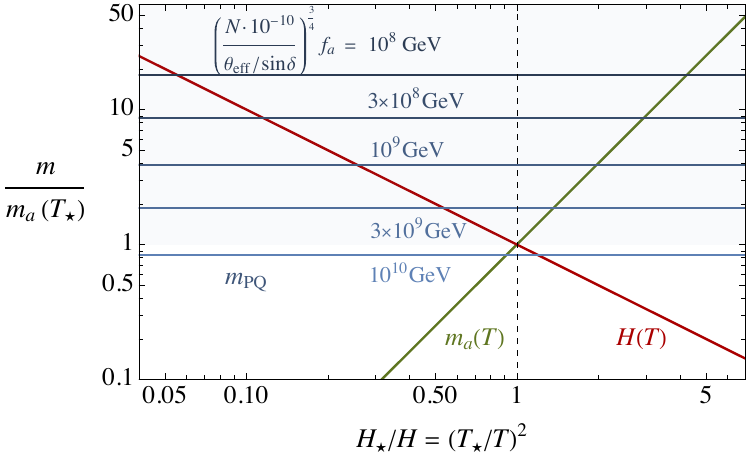}
    \caption{{\small Time evolution of the axion mass from QCD, $m_a$, (green) and the Hubble parameter (red). The blue lines show different values of the (temperature-independent) PQ-breaking mass $m_{\rm PQ}$, and can be interpreted as the $m_{\rm PQ}$ that saturates the maximal allowed PQ violation ($\theta_{\rm eff}=10^{-10}$) for different $f_a$%
    . For $f_a\lesssim 10^{10}\,\text{GeV}$ (blue region), the PQ-breaking mass is larger than the Hubble parameter $H(T_\star)=m_a(T_\star)\equiv H_\star$ when the QCD mass becomes cosmologically relevant: unstable domain walls then form already at $H=m_{\rm PQ}>H_\star$ (where the red and blue lines cross), destroying the string network before the QCD domain walls can even form at $H=H_\star$.%
    }
    }
    \label{fig:evolutionmass}
\end{figure}
Comparing with the nEDM bound in Eq.\,\eqref{eq:maxPQviolation}, we see that for $f_a \lesssim 10^{10}\,\mathrm{GeV}$ and $N$ not too large, $m_{\rm PQ}$ can be larger than $m_a(T_\star$), and thus the PQ-breaking term can become cosmologically relevant before the QCD term without spoiling the solution to the strong CP problem. This is evident from Fig.~\ref{fig:evolutionmass}, where we show the time-evolution of the different mass contributions and the Hubble parameter for varying $f_a$%
. Combining Eqs.\,\eqref{eq:theta-eff} and \eqref{eq:ma2soma2}, for generic $\theta_{\rm eff}$ (i.e., $m_{\rm PQ}$) and $\delta$ the condition %
for the PQ-breaking potential to dominate at $T_\star$ is%
\begin{equation}\label{eq:fatheta-lim}
    \frac{m^2_{\rm PQ}}{m^2_a(T_\star)}\simeq\frac{1}{N}\left[\frac{10^{10} {\rm GeV}}{f_a}\right]^{\frac43}\frac{\theta_{\rm eff}/\sin\delta}{1.4\cdot 10^{-10}}\gtrsim1  \ . %
\end{equation}
If Eq.\,\eqref{eq:fatheta-lim} is not satisfied (e.g., when $\theta_{\rm eff}$ lies well below the allowed bound), PQ breaking will still eventually become cosmologically relevant, but only after the QCD potential has already set in. In Fig.~\ref{fig:results} we show the parameter space in terms of $\theta_{\rm eff}$ and $f_a$. The two regimes, which we now discuss, are separated by $m_{\rm PQ}/m_a(T_\star) \simeq 1$, when the inequality of Eq.\,\eqref{eq:fatheta-lim} is saturated. The precise location of the transition depends on an order-one numerical factor that enters the right-hand-side of Eq.\,\eqref{eq:fatheta-lim}, which sets the critical $m_{\rm PQ}/m_a(T_\star)$. The black lines in Fig.~\ref{fig:results} show the result for this parameter equal to $0.32$ and $1.8$ for $N=2$, which, as discussed
below, bracket its plausible range. 
At fixed $\theta_{\rm eff}$, larger $N$ tightens the requirement in Eq.\,\eqref{eq:fatheta-lim}, lowering the maximum allowed $f_a$, and hence the black lines in Fig.~\ref{fig:results}, by a factor $(N/2)^{3/4}$.

\subsection{PQ breaking important before the QCD potential}\label{ss:PQbefore}

When Eq.\,\eqref{eq:fatheta-lim} is satisfied (below the uncertainty band given by the black lines, e.g., $f_a \lesssim 10^{10}\,\text{GeV}$ with maximal PQ violation), %
the PQ-breaking contribution becomes important for the string network before the QCD contribution does. Since $V_{\rm PQ}$ has trivial domain-wall number, once the Hubble radius exceeds the wall thickness $m_{\rm PQ}^{-1}$, i.e.\ at $H = m_{\rm PQ}$, a single domain wall of $V_{\rm PQ}$ forms attached to each string, located at $a(x)/v = \pi + \delta$.  As in the standard $N=1$ QCD case, this string-wall network is unstable and collapses rapidly. The QCD-generated potential becomes relevant only afterwards, at $H = m_a(T_\star)$. By then either no strings remain or the network is already heavily biased, and no stable $N>1$ domain-wall network can subsequently form, so the domain-wall problem is avoided (population-biased networks were shown to decay in \cite{Larsson:1996sp}). As discussed in Sec.\,\ref{ss:PQafter}, if the strings survive past this stage, the QCD domain walls formed at $T_\star$ will be destroyed only much later by the potential bias $V_{\rm PQ}$, after the axion mass has saturated.

To check these dynamics we simulated the string-wall system arising from Eqs.\,\eqref{eq:Lag} and \eqref{eq:QCDpotential}, including both $V_{\rm QCD}$ and $V_{\rm PQ}$, with $N=2$%
. Such simulations also allow us to extract the  minimum value of $m_{\rm PQ}/m_a(T_\star)$ for which the network decays before $V_{\rm QCD}$ generates stable DWs that would persist until $T_c$. We call this critical value $c_{\rm PQ}$, which is expected to be order-one. However, 
as detailed in App.~\ref{app:simulations},  the large hierarchy between the string thickness ($\sim m_r^{-1}$) and the DW width ($\sim m_{\rm PQ}^{-1}$ or $\sim m_a^{-1}$) makes a full simulation of the string and DW evolution hopeless: only values of $m_r$ with $\log_{\rm PQ}\equiv \log(m_r/m_{\rm PQ})\lesssim6$ are directly testable, whereas the physically relevant regime corresponds to $\log_{\rm PQ}\sim60-70$. As a result, extrapolation and caution are needed when interpreting the numerical value of $c_{\rm PQ}$ inferred from simulations. A conservative estimate of the importance of logarithmic corrections (see App.~\ref{app:decay-time}) leads to $c_{\rm PQ} \simeq 1.8$.

\begin{figure}
    \centering
    \includegraphics[width=0.55\linewidth]{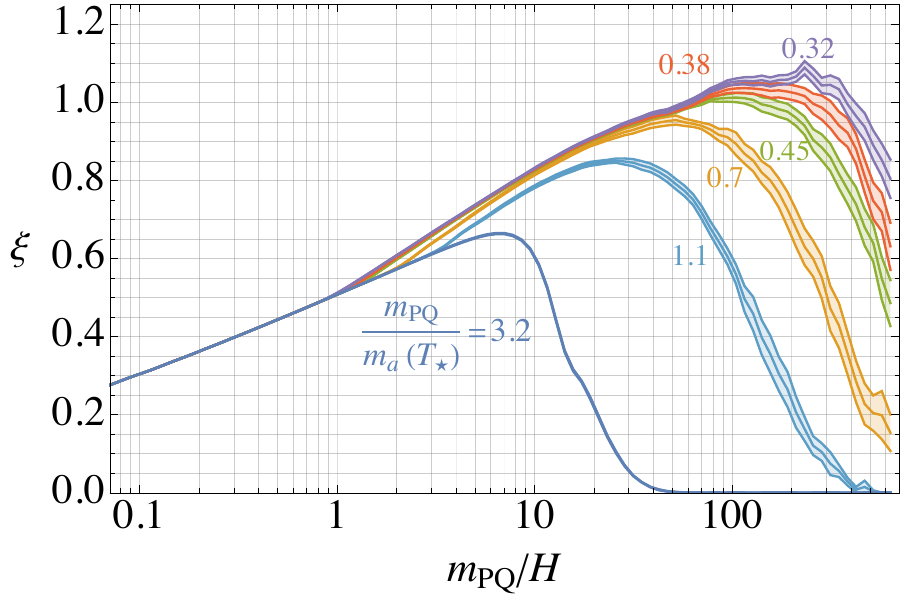}
    \caption{{\small Evolution of the number of strings per Hubble patch $\xi$ in the presence of the PQ-breaking and QCD potentials, for different PQ-violating masses $m_{\rm PQ}$ (related to $f_a$ and $\theta_{\rm eff}$ via Eq.\,\eqref{eq:fatheta-lim} and Fig.\,\ref{fig:evolutionmass}). We fixed $N=2$ and $\log(m_r/m_{\rm PQ})=4.5$. The scaling regime ends when $H\simeq m_{\rm PQ}$, and the unstable DWs from the PQ-violating potential make the string-wall network decay. %
    The network decays for $m_{\rm PQ}/m_a(T_\star)\gtrsim0.32$, with the decay being delayed as $m_{\rm PQ}/m_a(T_\star)$ decreases.
    }%
    }
    \label{fig:evolutionstrings}
\end{figure}

In Fig.~\ref{fig:evolutionstrings} we show the time evolution of the number of strings per Hubble patch $\xi$, from the last part of the scaling regime, where it grows logarithmically as $\xi\propto \log(m_r/H)$ (see e.g.~\cite{Gorghetto:2020qws}), through the period when $H\lesssim m_{\rm PQ}$ and $V_{\rm PQ}$ first, and $V_{\rm QCD}$ later, become relevant. In different lines we vary $m_{\rm PQ}/m_a(T_\star)$, but keep $\log_{\rm PQ}=4.5$ fixed (this is the largest value for which the DWs decay sufficiently early to keep finite-volume effects under control given our lattice sizes). At $H\simeq m_{\rm PQ}$ the scaling regime stops. The value $m_{\rm PQ}/m_a(T_\star)=3.2$ is large enough that the network fully collapses under the effect of $V_{\rm PQ}$ before $V_{\rm QCD}$ has any influence. For smaller $m_{\rm PQ}/m_a(T_\star)$, the effect of $V_{\rm QCD}$ becomes more substantial and the network takes longer to decay. Correspondingly, $\xi$ rises as the QCD potential starts to produce a possibly stable DW network with $N=2$. Even when the network has not fully collapsed before the QCD potential becomes dominant (e.g.\ for $m_{\rm PQ}/m_a(T_\star)=0.32$), the collapse nevertheless continues, and no stable DW network forms. This is because the string-wall network is already heavily biased from the onset of collapse that occurred at $H\simeq m_{\rm PQ}$, before the QCD potential came to dominate (see the axion field distribution in Fig.~\ref{fig:axion-distribution-low} in App.~\ref{app:field-distribution}). The results shown in Fig.~\ref{fig:evolutionstrings} suggest that $m_{\rm PQ}/m_a(T_\star)\gtrsim 0.32$ is sizable enough to ensure that the network decays, at least for the (rather moderate) logarithm value, $\log_{\rm PQ}=4.5$, explicitly simulated by us. %
Note that resolving the thickness of the QCD DWs on the lattice requires $m_a(T)$ to saturate unphysically at late times, at a value well below its zero-temperature one. To avoid introducing spurious effects from the PQ-breaking term, we set $m_{\rm PQ}$ to zero from that point. This  underestimates the effect of $V_{\rm PQ}$ and therefore overestimates $c_{\rm PQ}$, making our result conservative as an upper bound.%

The results in Fig.~\ref{fig:evolutionstrings} would point to $c_{\rm PQ} \simeq 0.32$. However, the tension of a global string such as the axion string, $\mu = \pi v^2 \log_{\rm PQ}$, has a logarithmic enhancement. At large $\log_{\rm PQ}$ such an enhancement might delay the effect of the DW tension, since the higher effective mass of the strings makes them more inert, therefore postponing the string-wall network's decay relative to the naive expectation at $H=m_{\rm PQ}$. This can be seen by estimating the acceleration of long parallel strings pulled together by a DW, which is of order $\sigma/\mu\propto 1/\log_{\rm PQ}$, with $\sigma \simeq 8 m_{\rm PQ} v^2$ the DW tension. %

We test this expectation with simulations in App.~\ref{app:decay-time} and find evidence that such a delayed decay does indeed occur, though we cannot extrapolate its precise dependence on $\log_{\rm PQ}$ to $\log_{\rm PQ}\gg5$. This introduces an uncertainty in the numerical value of $c_{\rm PQ}$, which we discuss in App.~\ref{app:decay-time}: conservatively estimating this dependence from the simulation data, we find that in the worst case the delay translates into a logarithmic scaling $c_{\rm PQ}\propto\log_{\rm PQ}^{2/3}$. At the physical point $\log_{\rm PQ}\simeq 60$, this gives an upper bound $c_{\rm PQ}\simeq 0.32\cdot (60/4.5)^{2/3}\simeq1.8$, which corresponds to the lower black line in Fig.~\ref{fig:results}.

Although we simulated $N=2$ only, we expect the smallest $m_{\rm PQ}/m_a(T_\star)$ for which the network decays early to depend on $N$ mildly, since the network is disrupted at $H=m_{\rm PQ}$, independently of $N$, which enters $V_{\rm QCD}$ only. Confirming this would however require explicit simulations.%

In this early PQ domination case, within the string scaling regime, the axions emitted from the strings contribute to the dark matter density, until at least $H=m_{\rm PQ}$. At this time, these axions are relativistic and carry an energy density at IR momenta (i.e., of order Hubble) $\rho_{\rm IR}\simeq 8\pi v^2 m_{\rm PQ}^2\xi_{\rm PQ}\log_{\rm PQ}$ (with $\xi_{\rm PQ}\simeq 0.24\log_{\rm PQ}$)~\cite{Gorghetto:2020qws}. 
The large enhancement factor, $\xi_{\rm PQ}\log_{\rm PQ}\simeq 10^3$, makes the energy density of these
(relativistic) axions much larger than the energy stored in the axion potential, which is bounded by
$m_{\rm PQ}^2v^2$ (and later by $m_a^2(T)f_a^2$). As a result they continue to redshift as radiation,
\begin{equation}
\rho_{\rm IR}(H)\simeq 8\pi v^2H^2\xi_{\rm PQ}\log_{\rm PQ} \, ,
\end{equation}
until $\rho_{\rm IR}\simeq V$. Only then do the axions become nonrelativistic. This delayed onset of nonrelativistic evolution suppresses their abundance relative to the case in which they become nonrelativistic already at $H=m_{\rm PQ}$. As shown in App.\,E.1 of~\cite{Gorghetto:2020qws}, the corresponding abundance nonetheless remains larger than that from preinflationary misalignment with $\theta_0=1$, $\Omega^{\theta_0=1}_{\rm mis}$, by a factor $Q\simeq (4\pi N^2 \xi_{\rm PQ}\log_{\rm PQ})^{7/12}$. The resulting string abundance is therefore 
\begin{equation}
\Omega_a^{\rm st}=Q\,\Omega^{\theta_0=1}_{\rm mis}\simeq (N f_a/10^{10}\,{\rm GeV})^{7/6}\,\Omega_{\rm DM} \, ,
\end{equation}
with $\Omega_{\rm DM}$ the measured dark matter density. Requiring that strings not overproduce dark matter thus bounds $f_a\lesssim 5\times 10^{9}\,{\rm GeV}$ for $N=2$ and $f_a\lesssim 2\times 10^{9}\,{\rm GeV}$ for $N=6$, as shown in blue in Fig.\,\ref{fig:results}.\footnote{In calculating $\Omega_a^{\rm st}$ we assume the axion emission spectrum from strings is IR dominated as suggested by \cite{Gorghetto:2020qws}. If the emission spectrum is instead scale invariant, the corresponding bound on $f_a$ is weakened by a factor of a few \cite{Buschmann:2021sdq}.}

As the PQ-breaking DWs decay, they produce an additional population of dark matter axions on top of $\Omega_a^{\rm st}$. Their abundance is unknown, but it is expected to be subdominant to $\Omega_a^{\rm st}$, since the DW energy density does not receive the logarithmic enhancement that boosts the energy stored in strings. Likewise, the misalignment contribution, with the initial angle set by the minimum of the PQ-breaking potential, $\theta_0=\delta$, is negligible compared to $\Omega_a^{\rm st}$. It is therefore plausible that $\Omega_a^{\rm st}$ provides a good approximation to the total dark matter abundance in this scenario. In the next subsection we consider the complementary regime in which a long-lived QCD domain-wall network forms before the PQ-breaking potential becomes important. The axions emitted by the decay of this long-lived network could instead easily overproduce the dark matter abundance.

We note that an important uncertainty arises from the presence of axions that have been produced by strings. Around $H_\star$, as mentioned, these axions contain a large energy density $\rho_{\rm IR}$ exceeding the energy density in the potential (specifically, this is the case at $\log_{\rm PQ}\gg 5$, a regime not accessible to simulations). This energy density might prevent the formation and decay of the PQ-breaking domain walls, 
delaying it until the time %
when $\rho_{\rm IR}\simeq V_{\rm PQ}$. In App.~\ref{app:waves} we show that if this is the case, the regime of early decay of strings shrinks considerably, as shown by the purple dashed line in Fig.~\ref{fig:results}. Future work clarifying the dynamics in this situation would be useful.

Note that, in the $N=1$ case, since the PQ-violating operators are %
cosmologically relevant only for $f_a \lesssim 10^{10}\,\rm GeV$, including their effect leaves unchanged the upper bound on the QCD axion decay constant from dark matter overproduction given in~\cite{Gorghetto:2020qws}. PQ-breaking operators would make the string network decay slightly earlier (at $H=m_{\rm PQ}$ rather than $H=H_\star$), but the axion dark matter abundance would not change significantly, since the axions during the relativistic redshift period at around $H=m_{\rm PQ}$ behave essentially like those emitted in the scaling regime. However, as recently noted in~\cite{Zantedeschi:2026iql}, only in the fine-tuned case $\delta \ll 1$ can the presence of PQ-violating operators relax this bound.

Finally, we comment on the dynamics expected for more general PQ-breaking potentials than considered so far. For a given $n$, the potential in Eq.\,\eqref{eq:Lag} would typically be accompanied by operators of the kind $|\phi|^{n-2}\phi^{3}/M_{p}^{\,n-3}+\dots$, with order-one coefficients and phases. For generic parameters, these operators lead to false vacua, separated from the true vacuum by a $\Delta V$ that is an order-one fraction of the potential's height, but do not spoil the uniqueness of the global minimum. In this case, DWs form at the time when $H\sim m_{\rm PQ}$, with $m_{\rm PQ}$ playing the role of the typical mass around the minima. The standard expectation is that such energy-biased DWs decay when $\Delta V\sim\sigma H$, with $\sigma\sim m_{\rm PQ} v^{2}$ the wall tension. As a result, the DWs are expected to decay when $H\sim m_{\rm PQ}$, immediately after they form. We therefore anticipate a phenomenology similar to the unit-charge case studied above, perhaps over a narrower range of parameters, although we stress that simulations would be needed to confirm this.

Another possibility is that the PQ-breaking potential might have multiple degenerate vacua, in which case it would, alone, lead to stable domain walls (this would occur if the leading operator has a definite charge greater than unity). Provided the total axion potential, including the QCD contribution, has a unique global minimum, early decay of the string network might also occur in this case. However, we leave a study of this case to future work.  The possibility of a PQ-breaking potential becoming cosmologically relevant before the QCD-induced one, in the scenario where the DWs induced by the PQ-breaking operator are stable, was already noted in Ref.~\cite{Barr:1992qq}.

\subsection{PQ breaking important after the QCD potential: long-lived DWs}\label{ss:PQafter}

If instead Eq.\,\eqref{eq:fatheta-lim} is not fulfilled (above the black lines in Fig.\,\ref{fig:results}), the PQ-breaking potential remains a small perturbation on the QCD one throughout. In this part of parameter space, the dynamics are as in the conventional picture: the QCD domain-wall network forming at $H=H_\star$ is long-lived, decaying only once the energy density of the bias potential $V_{\rm PQ}$ matches that of the QCD walls:
\begin{equation}\label{eq:Tdecay}
    2\mathcal{A}C_\sigma m_a f_a^2 H \simeq m_{\rm PQ}^2 v^2  \, ,
\end{equation}
with $\mathcal{A}%
$ the wall area per Hubble patch, and QCD wall tension $\sigma=C_\sigma m_a f_a^2$, $C_\sigma\simeq 8-9$.  Since, by assumption, the right-hand side of Eq.\,\eqref{eq:Tdecay} is much smaller than the wall energy density at $T_\star$, the decay condition is not satisfied when the QCD DW network first forms. For $T>T_c$, the QCD axion mass grows as $m_a\propto T^{-4}\propto H^{-2}$, so the left-hand side of Eq.\,\eqref{eq:Tdecay} increases with time, while the right-hand side remains constant. As a result, the decay condition can only be fulfilled once the QCD axion mass has reached its zero-temperature value. The domain-wall network therefore survives until a temperature $T_d<T_c$, determined by
\[
\frac{H(T_d)}{m_a(0)}\simeq \frac{N\theta_{\rm eff}}{2\mathcal{A}C_\sigma\sin\delta}\,.
\]
The energy stored in the domain walls is then released predominantly into nonrelativistic or mildly relativistic axions, providing an additional contribution to the axion dark matter relic abundance~\cite{Kawasaki:2014sqa,Gorghetto:2022ikz%
}. 
The corresponding density today can be estimated by redshifting nonrelativistically the energy density in Eq.\,\eqref{eq:Tdecay} from the time of DW decay,
\begin{equation} \label{eq:rhoa_w}
    \rho_a\simeq 2 \mathcal{A} \sigma H(T_d) \frac{g_{*s}(T_0)T_0^3}{g_{*s}(T_{d})T_{d}^3} \, ,
\end{equation}
which leads to 
\begin{equation}\label{eq:omega_aDW}
    \frac{\Omega_{\rm DW}^{N>1}}{\Omega_{\rm DM} } \simeq 0.85 \left[ \left(\frac{\mathcal{A}f_a}{5\cdot10^8\,{\rm GeV}}\right)^3\frac{10^{-10}}{N\theta_{\rm eff}/\sin\delta} \left(\frac{10}{g_{*,s}(T_d)}\right)^\frac12\right]^\frac12 \, .
\end{equation} 
Note that $\mathcal{A}$ encodes the area of all $N$ DWs, and thus is likely proportional to $N$. The estimate in Eq.\,\eqref{eq:omega_aDW} is crude and could receive large corrections, for instance from the exact decay temperature, which may differ from the $T_d$ calculated above, and the momentum of the axions emitted from DWs. Nevertheless, Eq.\,\eqref{eq:omega_aDW} matches parametrically the abundance found in simulations~\cite{Kawasaki:2014sqa}, and reproduces their $N=2$ result numerically if we choose $\mathcal{A}/N\simeq 0.6$. Given the large uncertainties, we refrain from a more detailed estimate of $\Omega_{\rm DW}^{N>1}$. We simply show in Fig.~\ref{fig:results} the dark matter overproduction bound (orange region) for $N=2$ and $\mathcal{A}/N=0.3-0.6$, which indicates the size of the uncertainty. %
The slope of this bound differs in the two regimes: for small $\theta_{\rm eff}/\sin\delta$ the dynamics are always dominated by late DW decay, whereas for large $\theta_{\rm eff}/\sin\delta$ the bound follows the transition between early and late DW decay.

Saturating the PQ-breaking operator at its maximum allowed value ($\theta_{\rm eff}=10^{-10}$), this long-lived DW scenario occurs only for $f_a \gtrsim 10^{10}\,\rm GeV$, where, as Eq.\,\eqref{eq:omega_aDW} and Fig.~\ref{fig:results} show, the DW network overproduces the observed dark matter abundance. For smaller PQ breaking (e.g., $\theta_{\rm eff} \lesssim 10^{-12}$ and $f_a \lesssim \text{few}\times 10^{8}\,\rm GeV$) the network decays in time and may avoid dark matter overproduction. However, such small $f_a$ values fall within the supernova (SN1987A) exclusion region for the QCD axion, shown in green in Fig.~\ref{fig:results}, see Refs.~\cite{PhysRevLett.60.1797,Raffelt:1987yt,Lella:2023bfb,Springmann:2024ret,Springmann:2024mjp} and~\cite{Fiorillo:2025gnd} for a recent analysis (the neutron-star cooling bound \cite{Buschmann:2021juv} is comparable). 
 A proper assessment of the viability of this scenario requires a careful extrapolation of numerical simulations of the string-wall system closer to the physical point, to validate the estimate in Eq.\,\eqref{eq:omega_aDW}, including both $\mathcal{A}$ and the network's decay time. In addition, at small $f_a$, thermal friction on strings from scattering of PQ fermions coupled to the axion may play an important role in the dynamics, including in the DFSZ model~\cite{Hook:2026grn}.

\section{Summary}
In this work we investigated the minimal scenario in which PQ-breaking operators %
are introduced to solve the axion DW problem, and identified a new, previously overlooked regime in which no long-lived DW network forms at all. Specifically, the parameter space $(\theta_{\rm eff},f_a)$ splits into two distinct regions. Below the black lines in Fig.\,\ref{fig:results}, PQ-breaking effects become important before the conventional stable QCD DW network can form; studying this regime both analytically and numerically, we find that the PQ-breaking potential annihilates the axion strings before the long-lived network forms. The black lines are shown for $N=2$ and scale as $N^{-3/4}$, so this scenario is viable also for e.g. $N=6$. Above the black lines, instead, the QCD axion potential becomes important first, and a long-lived DW network does form. This network eventually decays due to the PQ-breaking operators, but leaves behind a %
large relic abundance of axions.
This overproduces the observed dark matter abundance, except for small values of $f_a$, which are in borderline tension with supernova and neutron-star cooling bounds.

We conclude that a consistent $N>1$ QCD axion post-inflationary cosmology can only occupy the white region in Fig.\,\ref{fig:results}, specifically requiring
\begin{equation}\label{eq:eqsummary}
    f_a \lesssim \frac{2}{N}\cdot5\cdot 10^{9}\, \text{GeV} \qquad \text{or} \qquad  m_a \gtrsim  \frac{N}{2}\,{\rm meV}\, ,
\end{equation}
and sufficiently large $\theta_{\rm eff}$. The bound in Eq.\,\eqref{eq:eqsummary} follows from dark matter overproduction of axions by strings in the scaling regime. It is possible that an axion with mass saturating this bound yields the correct dark matter abundance. Combined with the supernova bound, our analysis points to a post-inflationary axion with mass in the $10^{-3}-10^{-2}$~eV range. The expected two-order-of-magnitude improvement in neutron EDM measurements~\cite{Alarcon:2022ero} will probe the natural part of this parameter space.
\vspace{.2cm}
\section*{Acknowledgments}%
We are grateful to Oriol Pujolas, Fabrizio Rompineve, Javi Serra,  Giovanni Villadoro and Edoardo Vitagliano for stimulating discussions. We thank Giovanni Villadoro for valuable feedback on a draft. The work of MG is supported by the Alexander von Humboldt foundation and has been partially funded by the Deutsche Forschungsgemeinschaft under Germany’s Excellence Strategy - EXC 2121 Quantum Universe - 390833306.  EH acknowledges the UK Research and Innovation Future Leader Fellowship MR/V024566/1. GP is supported by the Israel Science Foundation (ISF), Minerva, the NSF-BSF, and the European Research Council (ERC, DM-Dawn, Grant Agreement No. 101199868). SS acknowledges financial support from the Spanish Ministry of Science and Innovation (MICINN) through the Spanish State Research Agency, under Severo Ochoa Centres of Excellence Programme 2025-2029 (CEX2024001442-S). This work is also part of the R\&D\&i project PID2023-146686NB-C31, funded by MICIU/AEI/10.13039/501100011033/ and by ERDF/EU. IFAE is partially funded by the CERCA program of the Generalitat de Catalunya.

\newpage
\appendix 
\section{More details on early PQ breaking \label{app:estimates}}

Here we give more details on the early PQ-breaking regime and the derivation of Eqs.\,\eqref{eq:ma2soma2} and \eqref{eq:fatheta-lim}.

We work in a Friedmann-Robertson-Walker spacetime with metric $ds^2=dt^2-R^2dx^2$; during radiation domination the scale factor grows as $R\propto t^{1/2}$, so $H\equiv \dot{R}/R=1/(2t)$. %
At $T\gg T_c$, the equation of motion (EoM) for the axion field $\theta\equiv a/f_a$, fixing the radial mode to its vacuum expectation value, $r(x)=0$ are
\begin{equation}\label{eq:aEoM}
\ddot{\theta}+3H\dot{\theta}-R^{-2}\nabla^2\theta+m_a^2(T)\sin\theta +Nm_{\rm PQ}^2\sin(\theta/N-\delta)=0 \ ,
\end{equation}
where $m_a^2(T)\equiv \partial_a ^2V_{\rm QCD}(a=0)$.

\vspace{2mm}
\noindent
{\bf Onset of the QCD term.} In the absence of PQ violation ($m_{\rm PQ}=0$), the mass term $m_a(T)$ becomes cosmologically relevant once $H(T_\star)=m_a(T_\star)\equiv H_\star$, which defines the temperature $T_\star$. At this time the QCD domain walls form. Using the Friedmann equation $H^2(T)=g_{*}(T)\frac{\pi^2}{90}\frac{T^4}{M_p^2}$, this gives
\begin{align}\label{eq:Tstar} 
T_\star&=(m_a(0)f_a)^\frac12\left[\left(\frac{\Lambda}{(m_a(0)f_a)^\frac12}\right)^\frac{\alpha}{2}\frac{M_p}{f_a}\frac{3\sqrt{10}}{\pi g^{1/2}_{*}(T_\star)}\right]^\frac{2}{4+\alpha}\simeq 2.6 \, {\rm GeV} \left[\frac{10^{10} {\rm GeV}}{f_a}\left(\frac{60}{g_{*}(T_\star)}\right)^\frac12\right]^\frac16
\\
H_\star&=m_a(T_\star)=m_a(0)\left[\frac{m_a(0) M_p}{\Lambda^2}\frac{3\sqrt{10}}{\pi g^{1/2}_{*}(T_\star)}\right]^{-\frac{\alpha}{4+\alpha}} \simeq 1.2\cdot 10^{-5}\,m_a(0) \left[\frac{f_a}{10^{10} \,{\rm GeV}}\left(\frac{g_{*}(T_\star)}{60}\right)^\frac12\right]^\frac23\, ,\label{eq:Hstar}
\end{align}
where we used %
$m_a^2(T)=m_a^2(0)(\Lambda/T)^\alpha$, with %
$\Lambda\simeq 150$ MeV, and $\alpha\simeq 8$. Eq.\,\eqref{eq:Hstar} reproduces Eq.\,\eqref{eq:ma2soma2} in the main text. Note that $T_\star\gg T_c\simeq 150$\,MeV for all $f_a$ considered, consistent with the fact that the expression used for $m_a(T)$ is only valid for $T\gg T_c$.

\vspace{2mm}
\noindent
{\bf Duration of PQ breaking domination.} If $m_{\rm PQ}\gtrsim m_a(T_\star)$, the PQ-violating term begins to dominate at $H=m_{\rm PQ}$, inducing the formation of an unstable domain wall network. This, as discussed, either destroys the string network or biases it significantly. The QCD term regains dominance only later, at $H=H_1$, defined by $m_a(H_1)=m_{\rm PQ}$:
\begin{equation}\label{eq:mpqoH1}
    \frac{m_{\rm PQ}}{H_1}=\left(\frac{m_{\rm PQ}}{m_a(T_\star)}\right)^{1 + \frac{4}{\alpha}}\simeq 0.8\left(\frac{10^{10}\GeV}{f_a}\right)\left(\frac{\theta_{{\rm eff}}/(N\sin\delta)}{10^{-10}}\right)^\frac34 \, .
\end{equation}
This is the ratio between the initial and final values of the Hubble parameter over the period during which the PQ-violating term is dominant.

\section{Early destruction of $N>1$ networks in simulations}\label{app:simulations}

\subsection{Simulation details}\label{app:simulation-details}
To confirm the dynamics discussed in Sec.\,\ref{ss:PQbefore} and calculate the minimum value $c_{\rm PQ}$ of $m_{\rm PQ}/m_a(T_\star)$ that leads to the early destruction of the string network, we simulate the EoM from the Lagrangian in Eq.\,\eqref{eq:Lag} for $n=0$, and including the axion potential from QCD. %
The Lagrangian is
\begin{equation}\label{eq:Ltot}
    \mathcal{L}=|\partial_\mu \phi|^2-\frac{m_r^2}{2v^2}\left(|\phi|^2-\frac{v^2}{2}\right)^2+v^2\left[\frac{m_a^2}{N^2}\left(\frac{|\phi|}{v/\sqrt{2}}\right)^N\cos\left(\frac{Na}{v}\right)+m_{\rm PQ}^2\frac{|\phi|}{v/\sqrt{2}}\cos \left(\frac{a}{v}-\delta\right)\right] \, , 
\end{equation}
where $m_a=H_\star (H_\star/H)^{\alpha/4}$ grows with time. Generic random initial conditions lead to a network of axion strings that evolves in the scaling regime under the EoM of Eq.\,\eqref{eq:Ltot}. Domain walls appear once the axion potential becomes relevant.

As is standard for this system, we multiply the QCD potential in Eq.\,\eqref{eq:Ltot} by the factor $(\sqrt{2}|\phi|/v)^N$, which vanishes inside string cores (where $\phi=0$ and $a$ would otherwise be undefined) and approaches one outside them, leaving the dynamics unaffected away from the string core. To maximize the available evolution time, we study the so-called `fat' string system, in which the radial mode mass decreases in time as $m_r(R)=m_{r,0}(R_0/R)$, where $m_{r,0}$ is the mass at the initial time $R=R_0$, so that the comoving string core width $m_r^{-1}R$ remains constant in time.\footnote{The fat-string system has better numerical properties. In particular, since a longer interval of cosmic time separates two fixed values of $m_r/H$, the system converges to the string scaling solution more quickly.}

We solve the EoM on a discrete lattice of fixed comoving size, using grids of $N_x^3= 3072^3$ points and a standard finite-difference algorithm. In practice, we evolve the simplified EoM
\begin{equation}\label{eq:EoM}
    f''-\nabla^2f+\frac{m_r^2}{m_{r,0}^2}\frac12 f(|f|^2-R^2)-R^3\left[\frac{1}{N}\frac{m_a^2}{m_{r,0}^2}\left(\frac{f^*}{R}\right)^{N-1}+\frac{m_{\rm PQ}^2}{m_{r,0}^2}e^{i\delta}\right] =0\,,
\end{equation}
where we introduced the dimensionless field $f\equiv R\phi/(v/\sqrt{2})$ and conformal time $\tau=\int^t dt'/R(t')\propto\sqrt{t}\propto R$, used the identity $\cos(Na/v)=(1/2)(\phi^N+{\phi^*}^N)/|\phi|^N$, and derivatives are taken with respect to the dimensionless variables $m_{r,0}\tau$ and $m_{r,0}x$. We parametrize cosmic time using $\log(m_r/H)$ while studying the string scaling regime, and using $m_{\rm PQ}/H$ when studying the decay of the string-wall network. We aim to run until $m_{\rm PQ}/H\gg1$, so that the domain walls generated by PQ breaking have sufficient time to destroy the network.

We set the PQ-breaking mass scale $m_{\rm PQ}$ and the axion mass $m_a(T)$ appearing in Eq.\,\eqref{eq:EoM} by fixing, respectively, $\log_{\rm PQ}\equiv \log(m_r|_{H=m_{\rm PQ}}/m_{\rm PQ})$ and the ratio $m_{\rm PQ}/m_a(T_\star)$. The latter can be related to $f_a$ and $\theta_{\rm eff}$ via Eq.\,\eqref{eq:fatheta-lim}, while the former is physically of order $\log_{\rm PQ}\simeq 60$; however, as we discuss below, we can only test values up to $\log_{\rm PQ}\lesssim 6$. We fix $\alpha=8$ and $\delta=-1$.

The maximum value of $\log(m_r/H)$, and thus of $\log_{\rm PQ}=\log(m_r/m_{\rm PQ})$, reachable in a simulation is set by three requirements. First, the physical lattice spacing $\Delta$ must be small enough to resolve the string cores, $m_r\Delta\lesssim 1$ at all times. Second, the simulation volume must contain at least a few Hubble patches, $HL\gtrsim 1$, where $L$ is the physical box size. We take $m_r\Delta=1$ and $HL=1.5$, which we have checked introduces no systematic uncertainties~\cite{Gorghetto:2018myk,Gorghetto:2020qws,Gorghetto:2022ikz}; this constrains $\log_{\rm PQ}<\log(m_r/H)\lesssim 7.6$ for the grid sizes we use. In the main text, we take $\log_{\rm PQ}=4.5$ to be able to simulate a few \emph{e}-folds past $H=m_{\rm PQ}$.

Third, the ratio $m_a/m_r$ must remain small enough that the potential in Eq.\,\eqref{eq:Ltot} produces physical domain walls (across which the radial mode stays close to its vacuum expectation value~\cite{Fleury:2015aca}) rather than walls that interpolate over the top of the radial mode potential. In addition, if $m_a/m_r$ grows too large, the evolution becomes unphysical, e.g. string-antistring pairs connected by domain walls nucleate from the vacuum. Since $m_a$ grows quickly and soon reaches $m_r$, we enforce this bound by saturating $m_a$ at the critical value $m_a=m_r/\kappa$ (with $\kappa=6$~\cite{Moore:2001px}) once $m_a/m_r=1/\kappa$ is reached, after which $m_a$ decreases in time together with $m_r$. This condition is automatically sufficient to resolve both the PQ-breaking and QCD domain walls, whose widths $\sim m_{\rm PQ}^{-1}$ and $\sim m_a^{-1}$ both exceed $m_r^{-1}$.

Note that once $m_a$ saturates and starts decreasing, keeping $m_{\rm PQ}$ fixed could bias the results, since it could make an otherwise stable QCD domain wall network more unstable. We therefore set $m_{\rm PQ}$ to zero once the axion mass saturates, as mentioned in Sec.\,\ref{ss:PQbefore}; this underestimates the effect of the PQ-breaking potential.

In all simulations, the evolution begins at $R=R_0$, when the Hubble parameter equals $H=m_{r,0}=v/\sqrt{2}$, with initial conditions chosen to give a string configuration as close as possible to the scaling regime, following the procedure described in~\cite{Gorghetto:2018myk,Gorghetto:2020qws}. Throughout the evolution we track the number of strings per Hubble patch, $\xi(t)=\lim_{l\to\infty}\ell_{\rm tot}(l)\,t^2/l^3$, where $\ell_{\rm tot}(l)$ is the total string length within a volume $l^3$. Deviations of $\xi$ from its scaling-regime value $\xi(t)\simeq0.24 \log(m_r/H)$ serve as a proxy for the effect of the domain walls.

\subsection{Logarithmic dependence of the decay time} \label{app:decay-time}

Fig.~\ref{fig:evolutionstrings} in the main text shows the evolution of $\xi$ across the period when $H\simeq m_{\rm PQ}$, for different $m_{\rm PQ}/m_a(T_\star)$ and fixed $\log_{\rm PQ}=4.5$, averaged over $20$ simulations. This suggests that the PQ-breaking DWs form at $H\simeq m_{\rm PQ}$, and that $m_{\rm PQ}/m_a(T_\star)\simeq 0.32$ is sufficient to trigger the network's decay -- or at least to bias it enough, around $H\simeq m_{\rm PQ}$, that the decay eventually occurs. 

As mentioned, extrapolating this result to large $\log_{\rm PQ}$ requires understanding whether the effect of the PQ-violating operator (and the corresponding formation of DWs) is delayed at large $\log_{\rm PQ}$ relative to $H\simeq m_{\rm PQ}$. This may indeed be the case, since the acceleration of long parallel strings due to the attached DWs scales as $\alpha\simeq\sigma/\mu \simeq 3 m_{\rm PQ}/\log_{\rm PQ}$, where $\sigma \simeq 8 m_{\rm PQ} v^2$ and $\mu \simeq \pi v^2 \log_{\rm PQ}$. %
Let us for simplicity assume infinitely long parallel strings separated by initial distance $d_0 \simeq 1/(\xi(t_0)^{1/2}H(t_0))$. The strings move due to Hubble friction in the nonrelativistic limit with velocity
\begin{equation}
   \dot{v} + H v = \alpha.
\end{equation}
 At the same time, the separation between strings obeys
\begin{equation}
   \dot{d}=H d -2 v.
\end{equation}
Using $H=1/(2t)$, these coupled equations can be solved to give
\begin{equation} \label{eq:eomparallelstrings}
d(t)=
\frac{\sqrt{t}}{9\sqrt{t_0}}
\left[
9d_0
-8\alpha t^{3/2}\sqrt{t_0}
+12\alpha t_0^2\log\!\left(\frac{t}{t_0}\right)
+8\alpha t_0^2
\right]\,,
\end{equation}
where $t_0$ is the time at which the domain walls form. Parallel strings meet once $d(t_{\rm meet})=0$.  As a first approximation, neglecting the logarithmic term, we obtain
\begin{equation}
    t_{\rm meet}/t_0 = \frac{\left(8 + \frac{9 d_0}{\alpha t_0^2}\right)^{2/3}}{4}.
\end{equation}
For large values of $\log_{\rm PQ}$, one can expand for $9 d_0/(\alpha t_0^2) \gg 8$ to find
\begin{equation}
    t_{\rm meet}/t_0 \propto \left(\frac{d_0}{\alpha t_0^2}\right)^{2/3} \sim \left(\frac{\log_{\rm PQ}}{\sqrt{\xi(t_0)}}\right)^{2/3}\sim \log_{\rm PQ}^{1/3},
\end{equation}
where in the last step we used $\xi \propto \log_{\rm PQ}$. Instead of neglecting the Hubble expansion, one can also solve Eq.\,\eqref{eq:eomparallelstrings} numerically for different values of $\log_{\rm PQ}$, and then fit a scaling with $\log_{\rm PQ}$. The scaling does not differ significantly from the one without Hubble expansion.

\begin{figure}
    \centering
    \includegraphics[width=0.55\linewidth]{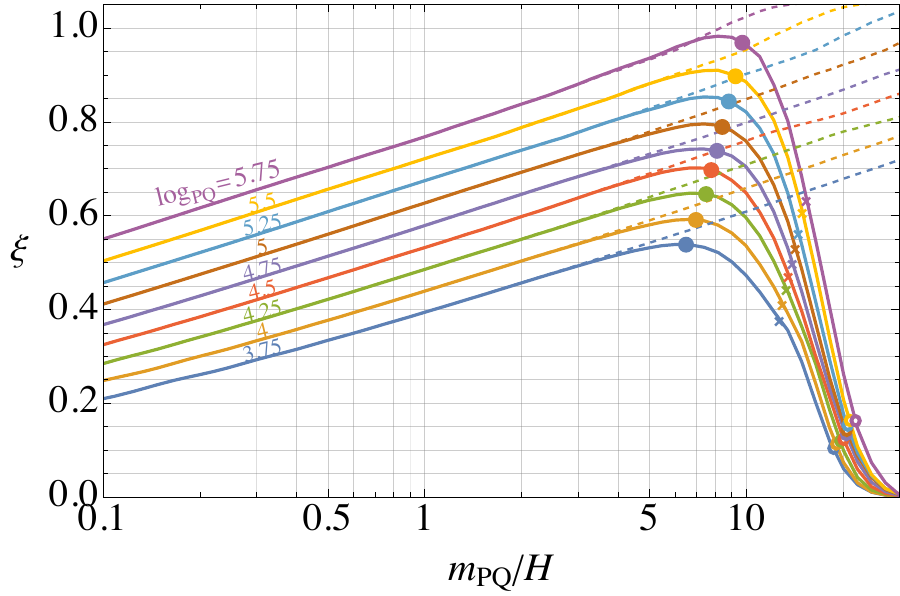} 
     \ \ \  \ \includegraphics[width=0.38\linewidth]{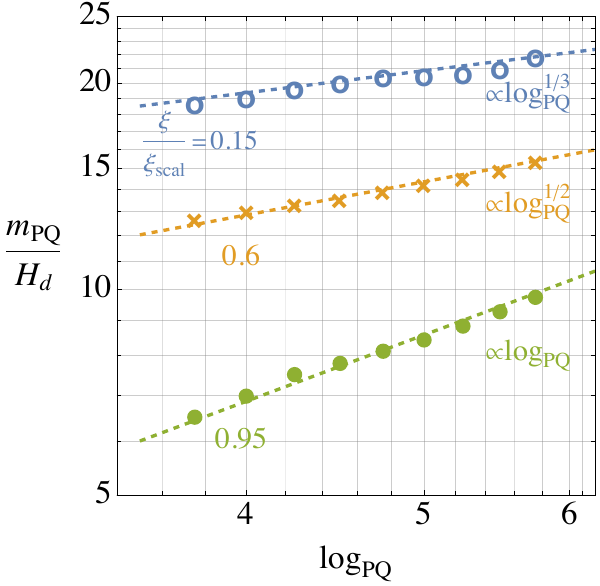} 
    \caption{{\small {\bf\emph{Left:}} Evolution of the string scaling parameter $\xi$ during the period when the PQ-breaking potential becomes relevant, for different choices of $\log_{\rm PQ}=\log(m_r/m_{\rm PQ})$ (solid lines), compared to the case without any axion potential (dashed lines). We set the QCD mass $m_a$ to zero. Filled circles, crosses, and empty circles mark the times $m_{\rm PQ}/H_d$ at which $\xi$ has decreased to $95\%$, $60\%$, and $15\%$ of its scaling-solution value (i.e. relative to the dashed lines). {\bf\emph{Right:}} Dependence of $m_{\rm PQ}/H_d$ on $\log_{\rm PQ}$.}} 
    \label{fig:simulation-log-dependence}
\end{figure}

To test this, in Fig.\,\ref{fig:simulation-log-dependence} (left) we show the evolution of the string-wall network including only the PQ-violating term in Eqs.\,(\ref{eq:Ltot}-\ref{eq:EoM}), i.e. setting $m_a=0$, for several choices of $\log_{\rm PQ}$ (solid lines). All choices share the same initial conditions, and results are averaged over 20 simulations. For comparison, we also show the scaling-solution evolution, i.e. without any axion potential (dashed lines). Unsurprisingly, larger $\log_{\rm PQ}$ yields a larger $\xi$ at $H=m_{\rm PQ}$ (proportional to $\log_{\rm PQ}$), reflecting the logarithmic growth of $\xi$ during the scaling regime. We then determine the Hubble time $H_d$ at which the network is affected by the PQ-violating term. We define $H_d$ as the Hubble time at which the string length has decayed to $\xi/\xi_{\rm scal}=95\%$, $60\%$, and $15\%$ of its scaling-solution value (filled circles, crosses, and empty circles, respectively, in Fig.\,\ref{fig:simulation-log-dependence}, left).

In Fig.\,\ref{fig:simulation-log-dependence} (right) we show this decay time (relative to $m_{\rm PQ}$) as a function of $\log_{\rm PQ}$, for each of the three choices of $\xi/\xi_{\rm scal}$. While a larger $\log_{\rm PQ}$ clearly leads to a larger decay time $m_{\rm PQ}/H_d$, this scaling is approximately proportional to $\log_{\rm PQ}$ only for the largest fractions, e.g. $\xi/\xi_{\rm scal}=95\%$. This suggests that the strings begin moving under the effect of the DWs increasingly later, proportionally to $\log_{\rm PQ}$: e.g. $m_{\rm PQ}/H_d\simeq(4-7)\log_{\rm PQ}$, where the range $4$--$7$ corresponds to a $1\%$--$5\%$ decrease in $\xi$.

By contrast, the time at which the string-wall system decays completely, or almost completely, depends more weakly on $\log_{\rm PQ}$: e.g. $m_{\rm PQ}/H_d\propto\log_{\rm PQ}^{1/3}$ for $\xi/\xi_{\rm scal}=15\%$, which agrees with our simplistic analytic estimate. %
In any case, since a network cannot decay before the strings begin to be affected by the DWs, for large $\log_{\rm PQ}$ all lines should converge to one asymptotic scaling. The simulations cannot access this behaviour and we choose the most conservative scaling, $m_{\rm PQ}/H_d\propto\log_{\rm PQ}$.

In this case, %
the minimum value $m_{\rm PQ}/m_a(T_\star)\simeq 0.32$ required for the string-wall system to decay grows with $\log_{\rm PQ}$, and must be revised accordingly. To this end, we use the following argument. If the strings only begin responding to the PQ-violating DWs a factor $\log_{\rm PQ}$ later in Hubble time, then the duration of the period of PQ-breaking domination over the QCD term is correspondingly reduced by the same factor $\log_{\rm PQ}$, relative to the estimate $m_{\rm PQ}/H_1$ in Eq.\,\eqref{eq:mpqoH1}. A value of $m_{\rm PQ}/H_1$ larger by a factor $\log_{\rm PQ}$ is therefore needed to compensate. Since $m_{\rm PQ}/m_a(T_\star)=(m_{\rm PQ}/H_1)^{2/3}$ from Eq.\,\eqref{eq:mpqoH1} with $\alpha=8$, this implies that the critical $m_{\rm PQ}/m_a(T_\star)$ must increase by a factor $\log_{\rm PQ}^{2/3}$, giving the logarithmic dependence $c_{\rm PQ}\propto\log_{\rm PQ}^{2/3}$ mentioned in Sec.\,\ref{ss:PQbefore}.

\begin{figure}
    \centering
    \includegraphics[width=0.49\linewidth]{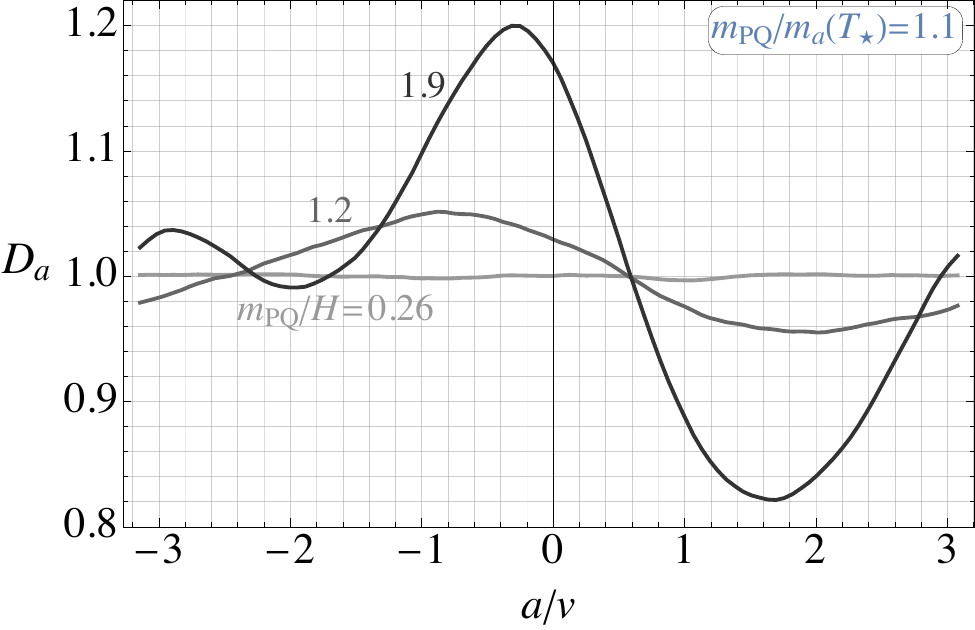} 
     \  \ \includegraphics[width=0.48\linewidth]{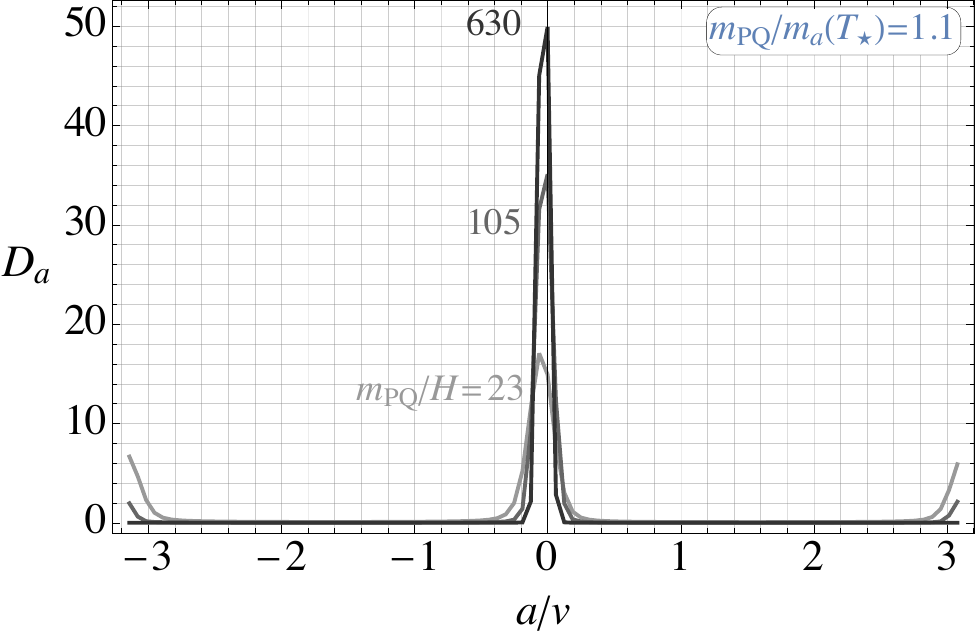} 
    \caption{%
    {\small Evolution of the axion field distribution $D_a$ for $m_{\rm PQ}/m_a(T_\star)=1.1$ (light blue line in Fig.\,\ref{fig:evolutionstrings}), during the early stages of PQ domination (left) and after the QCD potential becomes relevant (right).}
    } 
    \label{fig:axion-distribution-high}
\end{figure}

\subsection{Axion field distribution} \label{app:field-distribution}

\begin{figure}
    \centering
    \includegraphics[width=0.49\linewidth]{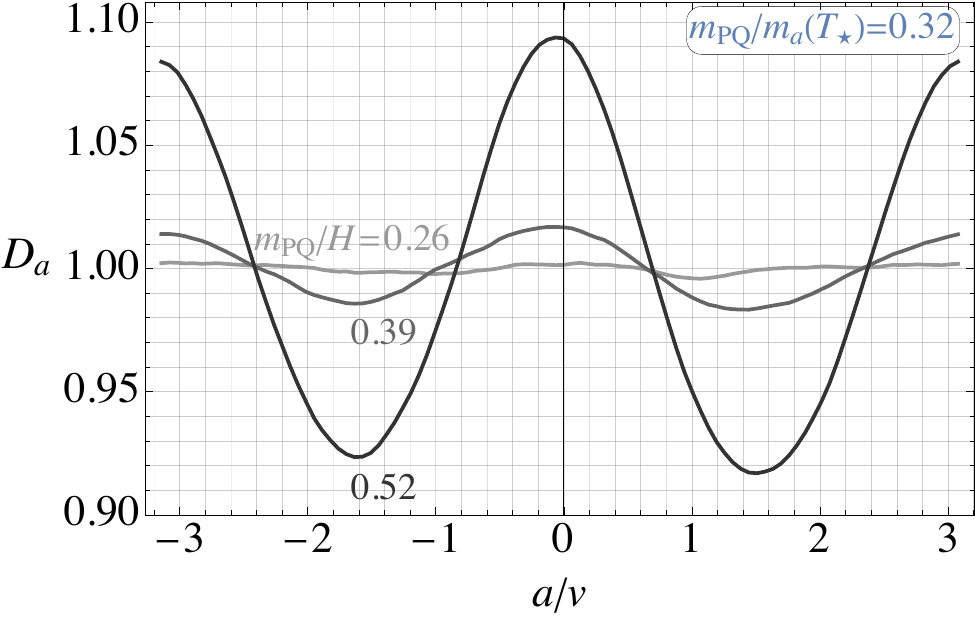} 
     \  \ \includegraphics[width=0.48\linewidth]{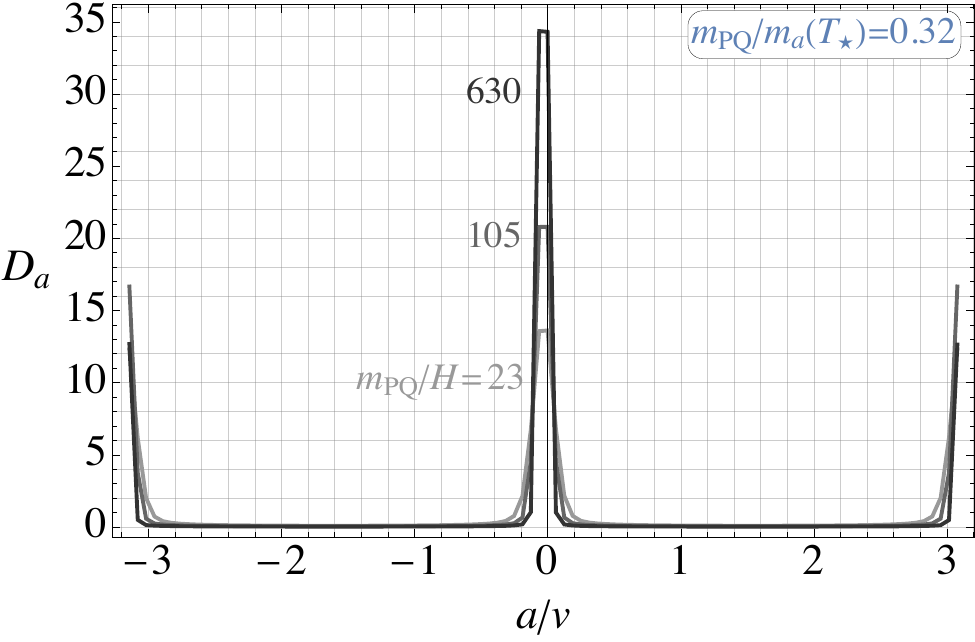} 
    \caption{{\small Evolution of the axion field distribution $D_a$ for $m_{\rm PQ}/m_a(T_\star)=0.32$ (purple line in Fig.\,\ref{fig:evolutionstrings}), during the early stages of PQ domination (left) and after the QCD potential becomes relevant (right).} }
    \label{fig:axion-distribution-low}
\end{figure}

Finally, we study the axion field distribution $D_a$ during DW collapse, defined as the fraction of spatial points where the axion field takes a given value in the interval $(-\pi v,\pi v]$, normalized to unity. Figs.~\ref{fig:axion-distribution-high} and \ref{fig:axion-distribution-low} show results for $m_{\rm PQ}/m_a(T_\star)=1.1$ and $0.32$, respectively (in the latter case, the PQ-breaking effect is the weakest among those shown in Fig.\,\ref{fig:evolutionstrings} of the main text, and the DWs decay the latest). The left panels show the evolution during the initial collapse, while the right panels show the network already decaying. At early times, when only strings are present ($m_{\rm PQ}/H\ll1$) and the axion potential is irrelevant, no value is preferred and the distribution is flat; see the left panels of Figs.~\ref{fig:axion-distribution-high} and \ref{fig:axion-distribution-low}, at $m_{\rm PQ}/H=0.26$.

The effect of $V_{\rm PQ}$ is to form a network of DWs that enhances the distribution around $a/v=\delta=-1$. For $m_{\rm PQ}/m_a(T_\star)=1.1$, a peak first develops at that position, at $m_{\rm PQ}/H\simeq1.2$; since the QCD potential is not yet dominant at this time, no peak is visible at $a/v=0$. By the later time $m_{\rm PQ}/H\simeq1.9$, the QCD potential has become important: the peak has shifted to $a/v\simeq0$, though it remains offset toward negative values of $a/v$ due to the PQ-breaking potential. Eventually $a/v=0$ comes to dominate the distribution entirely, at the expense of $a/v=\pi$ (Fig.~\ref{fig:axion-distribution-high}, right).

For weaker PQ breaking, $m_{\rm PQ}/m_a(T_\star)=0.32$, the picture differs: rather than being affected first by the PQ potential and only later by the QCD one, the distribution is distorted by both at approximately the same time (Fig.~\ref{fig:axion-distribution-low}, left). The PQ term is sufficient to slightly bias the distribution toward negative $a/v$, and this small bias is ultimately enough for $a/v=0$ to dominate over $a/v=\pi$ (Fig.~\ref{fig:axion-distribution-low}, right), driving the domain-wall network to decay. This is also evidence  that the network decays because of the PQ bias, rather than as a finite-volume artifact.

\section{Axion-wave obstruction of domain-wall decay}\label{app:waves}

There is the possibility that at the physical point where $\log(m_r/m_{\rm PQ})\simeq\log(m_r/m_a(T_\star))\simeq60\text{--}70$, the DWs attached to the strings might not form at all at either $H=m_{\rm PQ}$ (for $m_{\rm PQ}\gg m_a(T_\star)$) or $H=m_a(T_\star)$ (for $m_{\rm PQ}\ll m_a(T_\star)$). The reason is that, around those times, the Hamiltonian density of the axion field,
\begin{equation}\label{eq:Ham}
\mathcal{H}=\frac12 \dot a^2+\frac12 |\nabla a|^2 -m_a^2(T)\,f_a^2\cos(a/f_a) - m_{\rm PQ}^2 v^2 \cos\left(\frac{a}{v}-\delta\right) \, ,
\end{equation}
is not dominated by the axion potential (the last two terms in Eq.\,\eqref{eq:Ham}), but by the energy density of relativistic axions (or axion `waves') radiated during the preceding string scaling regime, encoded in the kinetic terms (the first two terms). Specifically, the potential $V$ is irrelevant at $H=m_{\rm PQ}$ and $H=m_a(T_\star)=H_\star$ because it is bounded from above by $m_{\rm PQ}^2v^2$ and $m_a^2(T)f_a^2$, respectively, while the IR energy density of these axion waves (i.e., at momenta of order Hubble) is $\rho_{\rm IR}(H=m_{\rm PQ})=8\mu\,\xi_{\rm PQ}\,m_{\rm PQ}^2$, with $\mu=\pi v^2\log_{\rm PQ}$. Thus, $\rho_{\rm IR}$ exceeds $V$ by a factor of at least $\xi_{\rm PQ}\log_{\rm PQ}\simeq\mathcal{O}(10^3)$; the same holds, analogously, at $H=H_\star$.

As long as $\rho_{\rm IR}\gg V$, including at $H=m_{\rm PQ}$ and $H=H_\star$, the potential energy is subdominant in the evolution of the system. Thus, the DWs attached to strings might %
not affect the network's evolution, even at times when $H<m_a$ or $H<m_{\rm PQ}$. The axion radiation energy density $\rho_{\rm IR}$ redshifts and becomes comparable to the potential only much later; it might be that only then the DWs %
cause the network to decay if $V_{\rm PQ}$ dominates over $V_{\rm QCD}$ at that time. This delayed onset of DW formation would imply a stronger condition for early DW decay from PQ domination than that given in Eq.\,\eqref{eq:fatheta-lim}. We stress, however, that these dynamics lie beyond the reach of direct simulation testing at present, and should be regarded as speculative.

\begin{figure}
    \centering
    \includegraphics[width=0.51\linewidth]{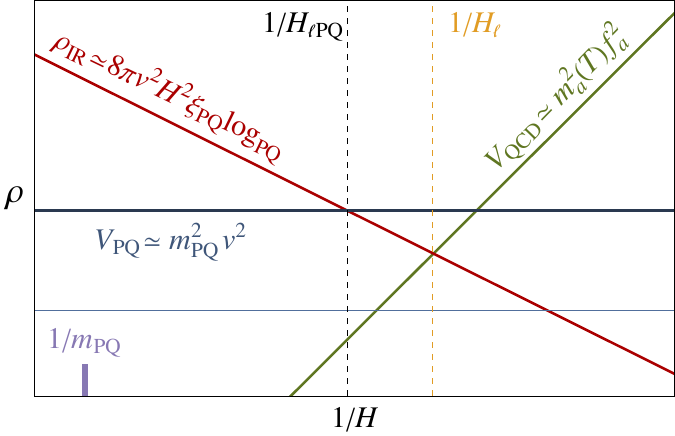}
    \caption{{\small {\small Sketch of the evolution of the different components of the Hamiltonian energy density in Eq.\,\eqref{eq:Ham}. At $H=m_{\rm PQ}$, the IR energy density of the axion waves $\rho_{\rm IR}$ exceeds both components of the potential, $V_{\rm PQ}$ and $V_{\rm QCD}$. For large PQ violation, $\rho_{\rm IR}$ reaches $V_{\rm PQ}$, at $H=H_{\ell{\rm PQ}}$, before it reaches $V_{\rm QCD}$. In this case the PQ-breaking operator would still trigger the early DW decay, even if axion waves had previously obstructed DW formation.}}}
    \label{fig:sketch}
\end{figure}

More specifically, the axions evolve according to the free Hamiltonian, i.e. $\rho_{\rm IR}(H)=\rho_{\rm IR}(H=m_{\rm PQ})(H/m_{\rm PQ})^2=8\pi v^2H^2\xi_{\rm PQ}\log_{\rm PQ}$, until the time when $\rho_{\rm IR}=c_V V$, where $c_V$ is an $\mathcal{O}(1)$ coefficient that can be computed from simulations. In the meantime, $V_{\rm QCD}\simeq m_a^2(T)f_a^2$ grows, while $V_{\rm PQ}\simeq m_{\rm PQ}^2v^2$ remains constant. See the red, blue, and green lines in Fig.~\ref{fig:sketch} for a sketch. We denote by $H_\ell$ and $H_{\ell{\rm PQ}}$ the Hubble parameters at which the equality $\rho_{\rm IR}=c_V V$ is satisfied, for $V=V_{\rm QCD}$ and $V=V_{\rm PQ}$, respectively; see the black and orange dashed lines.

If $H_{\ell{\rm PQ}}>H_\ell$, then once the axion potential starts affecting the axion dynamics, the PQ-breaking potential \emph{still} dominates over the QCD one, and its DWs (which by then would certainly have formed around the strings) have time to drive the decay of the $N>1$ string-wall network. This condition is satisfied, for example, by the thick blue line in Fig.~\ref{fig:sketch}, but not by the thin blue one. Note that $H_{\ell{\rm PQ}}>H_\ell$ is not automatically verified, since $V_{\rm QCD}$ grows rapidly in time due to the temperature dependence of the axion mass. Moreover, once $\rho_{\rm IR}=c_V V$, the dynamics of the axion waves becomes nonlinear, and the energy densities cease to follow the scalings shown in Fig.\,\ref{fig:sketch}.

The ratio $H_{\ell{\rm PQ}}/H_\ell$ can be computed following App.\,E.1 of Ref.~\cite{Gorghetto:2020qws}. That analysis considers $V=V_{\rm QCD}$ and provides the expression
\begin{equation}
z\equiv (m_a(H_\ell)/H_\star)^{1+6/\alpha}=\left[\frac{4\pi N^2\xi_{\rm PQ}\log_{\rm PQ}}{c_V}\left(1-\frac{2}{\alpha+4}\right)\log(\dots)\right]^{\frac{1}{2}\left(1+\frac{2}{\alpha+4}\right)} \, ,
\end{equation}
where the $\log(\dots)$ factor is subleading. Since $m_a(H_\ell)=H_\star(H_\star/H_\ell)^{\alpha/4}$, we also have $z=(H_\star/H_\ell)^{(6+\alpha)/4}$. Combining these two expressions for $z$ gives
\begin{align}\label{eq:HloHs}
    \frac{H_\ell}{H_\star} & = \left.\left[\frac{4\pi N^2\xi_{\rm PQ}\log_{\rm PQ}}{c_V}\left(1-\frac{2}{\alpha+4}\right)\log(\dots)\right]^{-\frac{2}{\alpha+4}}\right|_{\alpha=8}=\left[\frac{(6/5)c_V}{4\pi N^2 \xi_{\rm PQ}\log_{\rm PQ}\log(\dots)}\right]^\frac16 \, ,\\ 
    \frac{H_{\ell {\rm PQ}}}{m_{\rm PQ}} & =\left[\frac{c_V}{2\pi  \xi_{\rm PQ}\log_{\rm PQ}\log(\dots)}\right]^\frac12 \, .\label{eq:HLpqompq}
\end{align}
The second equation follows from the first by setting $\alpha=0$, and can equivalently be obtained directly from the definition $8\pi v^2 H_{\ell {\rm PQ}}^2\xi_{\rm PQ} \log_{\rm PQ}\simeq c_V m_{\rm PQ}^2 v^2$ modulo a $2\pi$ factor. Taking the ratio of Eq.\,\eqref{eq:HLpqompq} to Eq.\,\eqref{eq:HloHs},
\begin{equation}
    \frac{H_{\ell {\rm PQ}}}{H_{\ell}} =\frac{m_{\rm PQ}}{m_a(T_\star)} \left[\frac{(5/3)^{1/2} N c_V}{2\pi\log_{\rm PQ}\xi_{\rm PQ}\log(\dots)}\right]^\frac13 \simeq  \left[\frac{2}{N}\right]^\frac16\sqrt{\frac{\theta_{{\rm eff}}/\sin\delta}{10^{-10}}}\left[\frac{10^{8}\,\GeV}{f_a}\right]^\frac23 \, ,
\end{equation}
where, in the numerical estimate, we took $c_V=1$ and $\log(\dots)=1$, and expressed $m_{\rm PQ}/m_a(T_\star)$ using Eq.\,\eqref{eq:fatheta-lim}.  In Fig.\,\ref{fig:results}, the region with $H_{\ell{\rm PQ}}>H_{\ell}$ for $N=2$ lies below the purple dashed line. Indeed, if the effect of the DWs remains inactive until $H_{\ell{\rm PQ}}=H_\ell$ is reached, a substantial part of the parameter space for early destruction of the network is excluded.

\printbibliography

\end{document}